\documentstyle[12pt,preprint,aps]{revtex}
\title {Magnetophonon resonance in quantum wells in tilted field.\\
What is concealed behind its angular dependence?}
\author{{V. V. Afonin$^{a,b}$, V. L. Gurevich$^{a,b}$,
R. Laiho$^{a}$, M. O. Safonchik$^{a,b}$, M. A. Shakhov$^{a,b}$,\\ M.
L. Shubnikov$^{a,b}$}\\ {\it $^{a}$ Wihuri Physical Laboratory,
University of Turku, FIN-20014 Turku,
Finland}\\$^{b}${\em A. F. Ioffe Institute, 194021 Saint
Petersburg, Russia}}
\begin{document}
\maketitle
\begin{abstract}
Magnetophonon resonance in quantum wells in a tilted magnetic
field $\bf B$ is investigated. Measurements of the Hall
coefficient and correspondingly of the carrier concentration as
a functions of magnetic field and temperature are simultaneously
performed. It is shown that the experimental data can be
interpreted in terms of a great sensitivity of the effect to the
variation of the two dimensional carrier concentration $n_s$ in
a certain concentration interval. In other words, the observed
angular dependence of the MPR amplitudes is a manifestation of
dependence of $n_s$ on the magnitude of the magnetic field $B$.

\end{abstract}

\section{Introduction}

Magnetophonon resonance (MPR) in semiconductors is reached every
time when the limiting frequency of a longitudinal optic phonon
$\omega_0$ equals the cyclotron frequency of an electron,
$\Omega$, times some small integer, $\cal N$ (see
Refs.~\cite{GF,R}). Along with cyclotron resonance, it has
become one of the main instruments of semiconducting compound
spectroscopy.

The advances in semiconductor nano-fabrication in recent years have
made available nanostructures of a great crystalline perfection and
purity. The electrical conduction and some other transport phenomena
in such specimens has been a focus of numerous investigations, both
theoretical and experimental. In particular, the discovery of MPR in
the quantum wells took place in the pioneering paper by Tsui,
Englert, Cho and Gossard~\cite{TECG}. The most detailed experimental
investigation of MPR in quantum wells has been done by Nicholas and
co-workers ($\!$\cite{N} and the references therein). It has been shown
that there is a qualitative difference between MPR in 2D and 3D
structures.

In the 2D case MPR can exist only in a relatively narrow
interval of electron concentrations $n_s$. This has been
indicated in Ref.~\cite{N} and explained qualitatively in
Ref.~\cite{AGL1}. In a special group of
experiments~\cite{TECG,N} an angular dependence of MPR has been
investigated. As is well known, the 2D magnetoconductance,
including the MPR~\cite{AGL2}, at high magnetic fields $B$
should depend on the combination $B\cos\theta$ (see, for
example,~\cite{M}). Here $\theta$ is the angle between the
magnetic field $\bf B$ and the perpendicular to the plane of the
well. One can easily understand this using the following
classical analogy.  In the 2D case the curvature of an
electron's trajectory (in the course of electron's periodic
motion in the plane) can be considered as nonexistent in the
direction perpendicular to the plane because of the electron's
interaction with the walls of the well. This means that all the
physical quantities can depend only on the perpendicular
component of the field. In particular, the position of the $\cal
N$th MPR is given by \begin{equation} B_{\cal
N}(\theta)\cos\theta = B_{\cal N}(0) \label{1} \end{equation}
where $B_{\cal N}(\theta)$ is the position of the MPR maximum
for $\bf B$ directed at the angle $\theta$ to the perpendicular
to the plane of the well while $$B_{\cal
N}(0)={m\omega_0c/e{\cal N}}.$$ Here $\omega_0$ is the limiting
frequency of the optic phonons (we will not discriminate between
the frequencies $\omega_l$ and $\omega_t$ --- because of the
insufficient accuracy of our experiment) and $m$ is the
effective mass. Experimentally the angular dependence has been
investigated by Tsui et al.~\cite{TECG} and Brummel et
al.~\cite{B1}.  They have observed the angular dependence of the
amplitude of MPR maximum that appeared to be very sharp whereas
according to Eq. (\ref{1}) the amplitude of MPR maximum should
be independent of $\theta$ at all. That makes a drastic
disagreement between the experiment and theory. This means that
there is some feature in the system considered depending on the
total magnetic field $B$ rather than on the combination
$B\cos\theta$.

One of the main characteristics of the sample is the carrier
concentration $n_s$. It is usually implied that it depends
neither on the temperature nor on the magnetic field. Usually it
is really so at low temperatures where most experiments with
nanostructures are performed. However, the MPR experiments are
made at relatively high temperatures, the highest amplitudes in
GaAs being observed at $T$ about 180 K (they depend on $\cal N$
but only slightly). It is natural therefore to check the
temperature and magnetic field dependence of the concentration.
In order to control the electron concentration $n_s$, we have
performed observation of MPR along with measurements of the Hall
effect in 2D structures. Thus the purpose of the present paper
is the investigation of the MPR, simultaneous measurement of
magnetic field and temperature dependence of $n_s$ and
interpretation of the obtained data.

\section{Experimental results}
Three series of GaAs/Al$_x$Ga$_{1-x}$As quantum well samples
grown by molecular beam epitaxy were cut into a shape of a
typical Hall bar for observation of the Shubnikov-de Haas (SdH)
and the MPR oscillations. To avoid overheating of the sample
during the magnetic field pulse, we chose the measuring current
to be sufficiently small (of the order of 5 $\mu$A). The
measurements were carried out over the temperature interval of
4.2 --- 300 K in pulsed magnetic fields up to $B=40$ T with the
pulse duration of 8 ms. The main tool for collecting the data in
our pulsed field installation is the data acquisition card with
four fast independent 1$\mu$s, 12 bit digital channels having
128 Kb buffer memory each.

The measured signal had a smooth nonlinear component with the
amplitude much bigger than the amplitude of the investigated MPR
oscillation. To single out the oscillation and to get rid of the
high frequency noise we used the software package, based on the
approximation of the curve by the polynomial minimum squares
method with the Gaussian weight function. The method permits one
to process the signals properly, particularly at the edges of
the interval of magnetic field variation. However, it brings
about some distortion of the form of oscillation, especially for
the peaks near the edges (namely, the oscillation shifts towards
smaller fields while its amplitude goes down). Nevertheless, if
an edge of the interval is within the same phase of the MPR, the
distortion of the last peak should be also the same for all the
curves and the results can be compared. As under rotation of the
specimen the maxima shift towards bigger fields [see
Eq.~(\ref{1})] the maximal pulse field $B_{\rm max}$ should also
have the angular dependence $B_{\rm max}/\cos\theta$. All the
rest parameters used for the processing remained the same for
all the pulses.

Well developed SdH oscillations periodic in $1/B$ were observed at
$T$=4.2 K and used to determine the values of the low-temperature
carrier concentrations of the samples, namely $n_s$=2.2, 2.3 and 4.0
$\times 10^{11}\,\mbox{cm}^{-2}$. As pronounced MPR oscillations
require sufficient optical phonon population they are usually observable at
elevated temperatures. For this reason, we applied the Hall geometry to
investigate the dependence of the carrier concentration on the applied
field at fixed temperatures between 80 --- 300 K. Correspondingly, the
temperature dependence of $n_s$ was determined between 80 --- 300 K for
$B$ between 0.95 T and 27 T (see Figs. 1 and 2).

The MPR oscillations, also periodic in $1/B$, were recorded at
different temperatures in the range $T$=170 --- 230 K. In these
measurements the magnetic field was tilted at an angle $\theta$.
From the data the amplitude and the field positions of the $\cal
N$= 2, $\cal N$= 3 and $\cal N$= 4 MPR oscillation peaks were
analyzed (for $\cal N$= 2 and  $\cal N$= 3 they are given in Figs. 3
--- 5). Depending on the sample and the temperature, the accuracy of
the results varied between 2 --- 5 per cent.

The magnetic field interval 6 --- 15 T is important for our
purpose. At the same time, the Hall resistance $\rho_{xy}$
shows a good linear dependence on the magnetic field already at
1~T (see Fig. 1), and the resonance at $\cal N$=4 corresponding
to 5.62 T is reliably observed. This shows that for
$B\gtrsim6\,$T concentration variation as a function of magnetic
field $B$ is rather big and the deviation from the linear
dependence of $\rho_{xy}(B)$ cannot be explained by the
corrections proportional to $(\Omega\tau)^{-2}$, $1/\tau$ being
the collision frequency of the conduction electrons.

In the relevant magnetic field interval 6 --- 12 T the rate of the
electron concentration variation is temperature dependent (see Fig.~%
1). It is about 1.5 per cent per 1~T at 170~K and 3 per cent per 1~T
at 200~K. From the low temperatures up to 140~K there is no
noticeable magnetic field dependence of concentration. This behavior
fully correlates with the temperature dependence of the Hall
coefficient. One can see a rather strong temperature dependence up
from 140 K. Mark that when either the temperature or magnetic field
goes up the concentration $n_s$ also goes up.

\section{Discussion of the results}
As one can expect, the relative rate of concentration variation
with the temperature (see Fig. 2) is bigger than with magnetic
field. The latter, however, is by a factor of 5~---~6 bigger as
compared with the concentration variation due to the spin
magnetic moment of the free electrons. (The well-known
corrections to the $g$-factor of electrons in quantum
wells~\cite{I} can be disregarded as they also depend only on
$B\cos\theta$~\cite{K}). This makes one think of the electrons
that tunnel from discrete levels into a quantum well.  Usually
the electron levels go down with the magnetic field $B$.
Indeed, as $B$ goes up the electron wave functions become nearer
to the nuclei and therefore their binding becomes more tight or,
in other words, the absolute value of the electron binding energy
goes up. As a result, the level goes down as well as the
electron concentration $n_s$ in the well. One of the possible
mechanisms where the absolute value of the electron binding
energy goes up with the magnetic field can be described as
follows. If one has a negative ion where the electron is bound
to the atom by dipole forces the magnetic field compression of
the electron wave function should decrease the interaction with
the atom. As a result, the probability that the electron will
not be bound at elevated temperatures enhances. This means that
the electron levels go up with the magnetic field and $n_s$
behaves in the same manner. Such ions could be either in the
cover layer of the structure or in the interface.

In the papers by Nicholas {\em et al.}~\cite{N} a very strong
dependence of the MPR amplitude was demonstrated in the region of high
electron concentrations. The MPR amplitude started to go down at
$n_s$=10$^{11}$\,cm$^{-2}$. At $n_s$=3$\cdot$10$^{11}$\,cm$^{-2}$
the decrease became extremely strong. The amplitude decreased by a
factor 12 under the increase of concentration from
$n_s$=3$\cdot$10$^{11}$\,cm$^{-2}$ to
$n_s$=5.5$\cdot$10$^{11}$\,cm$^{-2}$.

This behavior is surprising. Indeed, the usual estimate of the
relative role of electron-electron (e-e) interaction as compared
with the kinetic energy of electrons gives $$
e^2n^{1/2}/\varepsilon k_{\rm B}T. $$ This quantity is of the
order of 1/4 at $n_s$=10$^{11}$\,cm$^{-2}$ and $T=200$\,K, i.e.
it seems that one can neglect the e-e interaction. The
degeneracy parameter becomes of the order of 1 at
$n_s\approx3\cdot$10$^{11}$\,cm$^{-2}$~\cite{N}.  However
usually the onset of degeneracy changes the effect by something
like a factor 2, or so, whereas a much more substantial
variation with the electron concentration was actually observed.

The physics of such a behavior has been described in
Ref.~\cite{AGL1}. Qualitatively it can be interpreted in the
following way. Usually one interprets MPR as a result of electron
transitions between two Landau levels. However another, a less direct
approach is also possible. One can treat MPR as enhancement of
interaction of a pair of electrons due to exchange of an optic phonon
(a pole of the scattering amplitude). One should, however, take into
consideration that, apart from the interaction due to exchange of a
phonon, the electrons have also a direct Coulomb interaction. The sum
of these two interactions can be described by a potential~\cite{GLF}
\begin{equation}
V={2\pi e^2\over q\varepsilon(\omega)}
\label{2}
\end{equation}
where
\begin{equation}
\varepsilon(\omega)=
\varepsilon_{\infty} \frac{\omega _{l}^{2}-(\omega+i\Gamma)^{2}}
{\omega_{t}^{2}-(\omega+i\Gamma)^{2}}.
\label{3}
\end{equation}
Here $\varepsilon_{\infty}$ is the lattice dielectric susceptibility
for $\omega\rightarrow\infty$, $\omega_l$ ($\omega_t$) are the
limiting frequencies of the longitudinal (transverse) optic phonons
while $\Gamma$ is the phonon damping due to the phonon anharmonicity.

Equation (\ref{3}) describes the direct interaction between two
electrons. One should, however, also allow for the indirect
interaction where the first electron interacts with the second
one and this, in its turn, interacts with the next electron.
Taking this in consideration, one should take into account the
following two points. First, one can consider any electron as
the next one. This will give the factor $n_s$. Second, the
interaction we are discussing is of a resonant nature. In the 2D
case the electron spectrum, unlike the 3D case, has no component
of the quasimomentum along the magnetic field. As a result, the
characteristic time of e-e interaction is not $m/\hbar q_z^2$ as
in 3D case but is determined by $1/\Gamma_e$ where $\Gamma_e$ is
the electron damping. This means that the electrons will be in
resonance during the time of the order of $1/\Gamma_e$. As a
result, we get for this interaction Eq. (\ref{2}) with an extra
factor~\cite{AGL2} $$ {2\pi e^2\over
q\varepsilon(\omega)}\cdot{n_s\over \hbar(\omega-{\cal N}\Omega
\cos\theta+i\Gamma_e)} .
$$
The whole expression is dimensionless.

Now we should take into account that this interaction may take place
1,2,3,$\dots$ times. As a result, the full interaction is
\begin{equation}
V_{\rm full}={2\pi e^2\over q\varepsilon
(\omega)}
\left[1-{n_s\over\hbar(\omega-
{\cal N}\Omega\cos\theta+i\Gamma_e)}\cdot
{2\pi e^2\over q\varepsilon(\omega)}\right]^{-1} .
\label{4}
\end{equation}
The interaction becomes very strong provided the expression in the
square brackets vanishes. In fact this condition means the existence
of electron transitions between Landau levels due to the
interaction with the mixed electron-phonon mode.

We will solve the equation
\begin{equation}
1-{n_s\over\hbar(\omega-
{\cal N}\Omega\cos\theta+i\Gamma_e)}\cdot
{2\pi e^2\over q\varepsilon(\omega)}=0 .
\label{5}
\end{equation}
by iterations considering the dampings as relatively small.  In the
lowest approximation we have two solutions we are
interested in
$$
\omega={\cal N}\Omega\cos\theta,\quad\mbox{and}\quad\omega=\omega_t .
$$
It means that
\begin{equation}
\omega_t={\cal N}\Omega\cos\theta.
\label{6}
\end{equation}
This condition determines the MPR peak positions. The next,
imaginary, approximation determines the width of the $\cal N$th
MPR peak
\begin{equation}
\Gamma_{\cal N}=\Gamma_{e}+{n_s\over n_{\rm up}}\Omega\cos\theta.
\label{7a}
\end{equation}
where
$$
n_{\rm up}={\varepsilon_{\infty}\hbar\Omega
\cos\theta(\omega_l-\omega_t)q\over2\pi
e^2\Gamma}.
$$
$\Gamma_{\cal N}$ should be smaller than the
spacing between the Landau levels. For large concentrations
$n_s$, in Eq.~(\ref{7a}) the last term is predominant. This gives
the condition
\begin{equation}
n_s/n_{\rm up}\ll1 .
\label{7}
\end{equation}

Here we have assumed that $$ (\omega_l-\omega_t)/\omega_l\ll1.  $$
and will neglect the terms proportional to this small parameter as
compared to 1. For the estimates we will take
$q=q_T\equiv\hbar^{-1}\sqrt{2mk_{\rm B}T}$.

When the parameter $n_s/n_{\rm up}$ is of the order of 1 the MPR
peaks begin to overlap and at $n_s/n_{\rm up}>1$ the MPR amplitude
should rapidly go down. However, if one follows the MPR maximum
whereas $n_s$ does not depend of the magnetic field $B$ the parameter
$n_s/n_{\rm up}$ does not change [see Eq. (\ref{6})] and, as a
result, the height of the MPR maximum remains constant. In the region
$n_s/n_{\rm up}>1$ even small variation of $n_s$ due to the variation
of $B$ may result in a strong variation of the height of the MPR
maximum. In case the height of the MPR peak depends only of the
parameter $n_s/n_{\rm up}$ the height would remain the same
irrespective of the way we change this parameter (it can be changed
either by magnetic field variation or by doping).

Let us check as to whether a relatively small concentration variation
due to a small magnetic field variation
\begin{equation}
\Delta B_{\cal N}(\theta)=B_{\cal N}(0)\left({1\over\cos\theta}-1\right)
\label{8}
\end{equation}
is sufficient to explain the decrease of the height of the MPR
maximum due to the tilting of the field $\bf B$ by the angle
$\theta$.  According to Ref.~\cite{N}, in the relevant interval of
electron concentrations under variation of $n_s$ by 45 per cent, i.e.
from 3$\cdot$10$^{11}$\,cm$^{-2}$ to 5.5$\cdot$10$^{11}$\,cm$^{-2}$
the amplitude of the maximum has been decreased by a factor of 12.
This means a doubling of the amplitude due to decrease of the
concentration by 2 per cent. The data given in the present paper are
obtained on the samples with the carrier concentrations in this
region. In Fig. 3 the amplitude of the second maximum at 170 K has
decreased by a factor 2 at the angle $\theta_{1/2}$=25$^{\rm o}$. The
variation of magnetic field is $\Delta B=1.13\,$T.  In the
perpendicular $\bf B$ the maximum is at 11.25 T. The variation of
electron concentration at this temperature is 1.5 per cent for 1 T
(Fig. 1). Thus the increase of the concentration is 1.7 per cent at
the angle of tilting 25$^{\circ}$. The agreement may be
considered as reasonable.

The rate of concentration variation as a function of magnetic field
goes up with the temperature (see Fig. 1). If the decrease of the MPR
amplitude is determined by growth of the concentration this
should enhance the sharpness of the angular dependence of the MPR
amplitude with the temperature. In fact this behavior has been
observed in our experiment (see Figs. 3, 4, 5). Indeed, in Fig. 4 at
T=190 K for $\cal N$=2, $\theta_{1/2}$=17$^{\circ}$. This corresponds to
a smaller variation of the field, $\Delta B=0.6\,$T. As, however, the
rate of concentration variation with $B$ goes up with higher
temperatures (it is of the order of three per cent per 1 T --- see
Fig. 1) one has in fact the same variation of the concentration
$\Delta n_s/n_s=1.8\,\%$. With our accuracy this coincides with the
drop of the MPR amplitude under the variation of the carrier
concentration in the perpendicular magnetic field $\bf B$ --- see
Ref.~\cite{N}. One has a decrease of the maximum by a factor 2 for
enhancement of the concentration $n_s$ by 2 per cent.

We can offer the following direct experimental proof that the
considered effect depends on the variation of the electron
concentration $n_s$ in the magnetic field. In the same sample, for the
same variation of the MPR amplitude it
is necessary that the variation of the concentration $n_s$ under
rotation of the sample should be the same for different values
of magnetic field.  In other words,
$$
\Delta n_s=\Delta B_{\cal
N}(\theta){\partial n_s\over\partial B}
$$
should be $\cal
N$-independent.  According to Fig.  1, in the interval of the field
variation 4 --- 12\,T the concentration is with our accuracy a linear
function of the field $B$. In other words, for $\cal N$=2 and 3,
$\,\Delta B_{\cal N}\!\left(\theta^{({\cal N})}_{1/2}\right)$ should be
$\cal N$-independent. Then, according to Eq.~(\ref{8}), we have for
example for $\cal N$=2 and 3 \begin{equation}
\cos\theta^{({3})}_{1/2}={B_{3}(0)\over B_{3}(0)
+B_{2}(0)\left(1\left/\cos
\theta\!\textstyle{(2)\atop{1/2}}\right.-1\right)} .
\label{9}
\end{equation}

The sharper is the peak, the more sensitive is Eq.~(\ref{9}) to the
variation of the angles $\theta_{1/2}$. As the peaks become more narrow
with the temperature, we have chosen $T$=230\,K. Then for $\cal N$=2 we
have $\theta_{1/2}$=12\,$^\circ$. As $B_2(0)=11.25\,$T, $B_3(0)=7.5\,$T
we get $\theta^{(3)}_{1/2}$=15\,$^\circ$ that is in a good agreement
with the experimental value (see Fig. 5).

\section{Conclusion}

In summary, we have investigated the magnetophonon resonance in
a tilted magnetic field measuring also the 2D electron
concentration of the same samples. Analyzing the experimental
data we have arrived at the following conclusions. The sharp
angular dependence of the MPR maxima on the magnetic field is a
manifestation of a very sharp concentration dependence of the
MPR amplitude in the perpendicular magnetic field.  The reason
as to why the 2D concentration of the carriers can enhance is,
as we understand, the following. At the tilting of the magnetic
field the MPR maximum is shifted towards the strong magnetic
fields \{see Eq.~(\ref{1}) and Fig. 6 that agrees with the data
of Ref.~\cite{N}\}.  The shift is comparatively small (of the
order of 1 T) and at high temperatures brings about
comparatively small concentration variation (of the order of
several \%). However, due to a very sharp concentration
dependence of the MPR amplitude, this is sufficient for a
decrease of the amplitude by several times in the relevant
concentration interval.

It would be very interesting to investigate in future the MPR in
quantum wells of various compositions. It is also desirable to
make a systematic investigation of the MPR in nanostructures of
different forms, such as quantum wires (see, for instance
Ref.~\cite{PG}) as well as to take into consideration the polaron
effect (cf with Ref.~\cite{P}). And of course it is important to
understand the behavior of the electron concentration $n_s$ as a
function of $T$, $B$ and $\theta$ in quantum wells of different
compositions, doping and dimensions.

Thus the principal conclusion of the paper can be formulated as
follows. The angular dependence of the MPR amplitudes as well as the
decrease of the resonance widths with the temperature is a
manifestation of dependence of $n_s$ on the total magnetic field $B$
(observed explicitly in the present paper). This statement
permits to relate three seemingly different groups of
experiments performed in different laboratories.
\begin{enumerate}
\item A sharp decrease of the MPR amplitude in perpendicular
magnetic field as a function of growing $n_s$ with a steep
angular dependence of the MPR amplitude under the tilting of the
magnetic field.

\item A narrowing of the angular dependence of the MPR peaks
as a function of rising temperature with the enhancement of the rate 
of variation of $n_s$ as a function of $B$.

\item The characteristic width of the MPR for different $\cal N$
with the rate of variation of $n_s$ as a function of $B$.
\end{enumerate}

We wish to emphasize that the dependence $n_s(B,T)$ has not been
an adjustable function. Rather it has been extracted from the
Hall effect measured on the same samples.

\acknowledgements

V. V. A., V. L. G., M. O. S., M. A. Sh. and M. L. Sh.  are grateful
to the Wihuri Foundation and to the Academy of Finland (project
79543) for a partial support of this work and to the Turku University
for hospitality. V. V. A. and V. L. G.  also acknowledge a partial
support for this work by the Russian National Fund of Fundamental
Research (Grant 00-15-96748).

\newpage
\bigskip
\centerline{\large FIGURE CAPTIONS}

\bigskip
1. Variation of 2D electron concentration $n_s$ as a function of
magnetic field $B$ for various temperatures.

2. Variation of 2D electron concentration $n_s$ as a function of
temperature $T$.

3. Angular dependence of the height of MPR maximum, 
$n_s=$4.0$\cdot10^{11}$cm$^{-2}$, T=170 K.

4. Angular dependence of the height of MPR maximum, 
$n_s=$2.2$\cdot10^{11}$cm$^{-2}$, T=190 K.

5. Angular dependence of the height of MPR maximum, 
$n_s=$4.0$\cdot10^{11}$cm$^{-2}$, T=230 K.

6. MPR maximum position as a function of the angle of tilting.
The broken line corresponds to the dependence
$B_{2}(0)/\cos\theta$ for $n_s=$4.0$\cdot10^{11}$cm$^{-2}$.

\bigskip

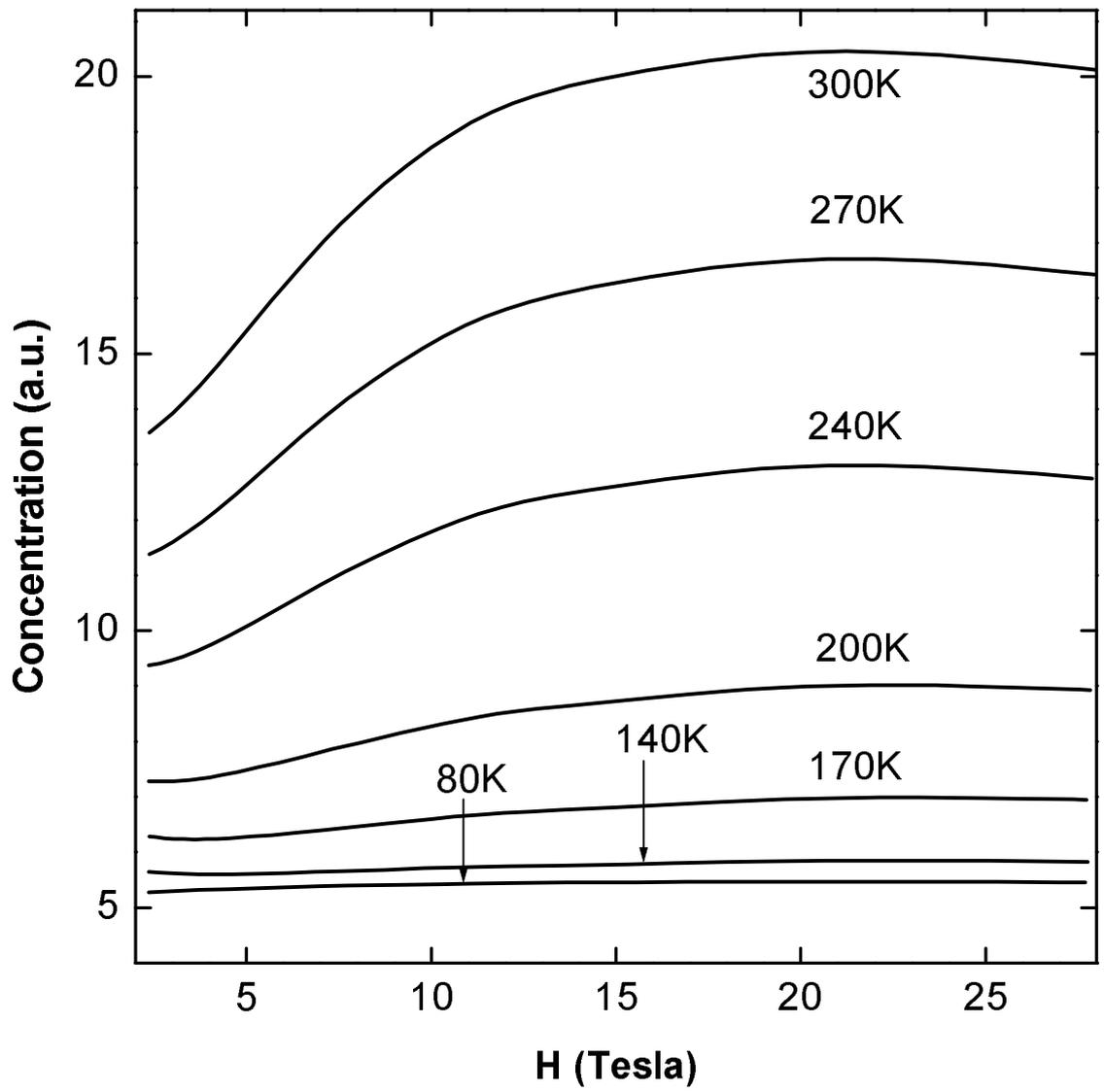
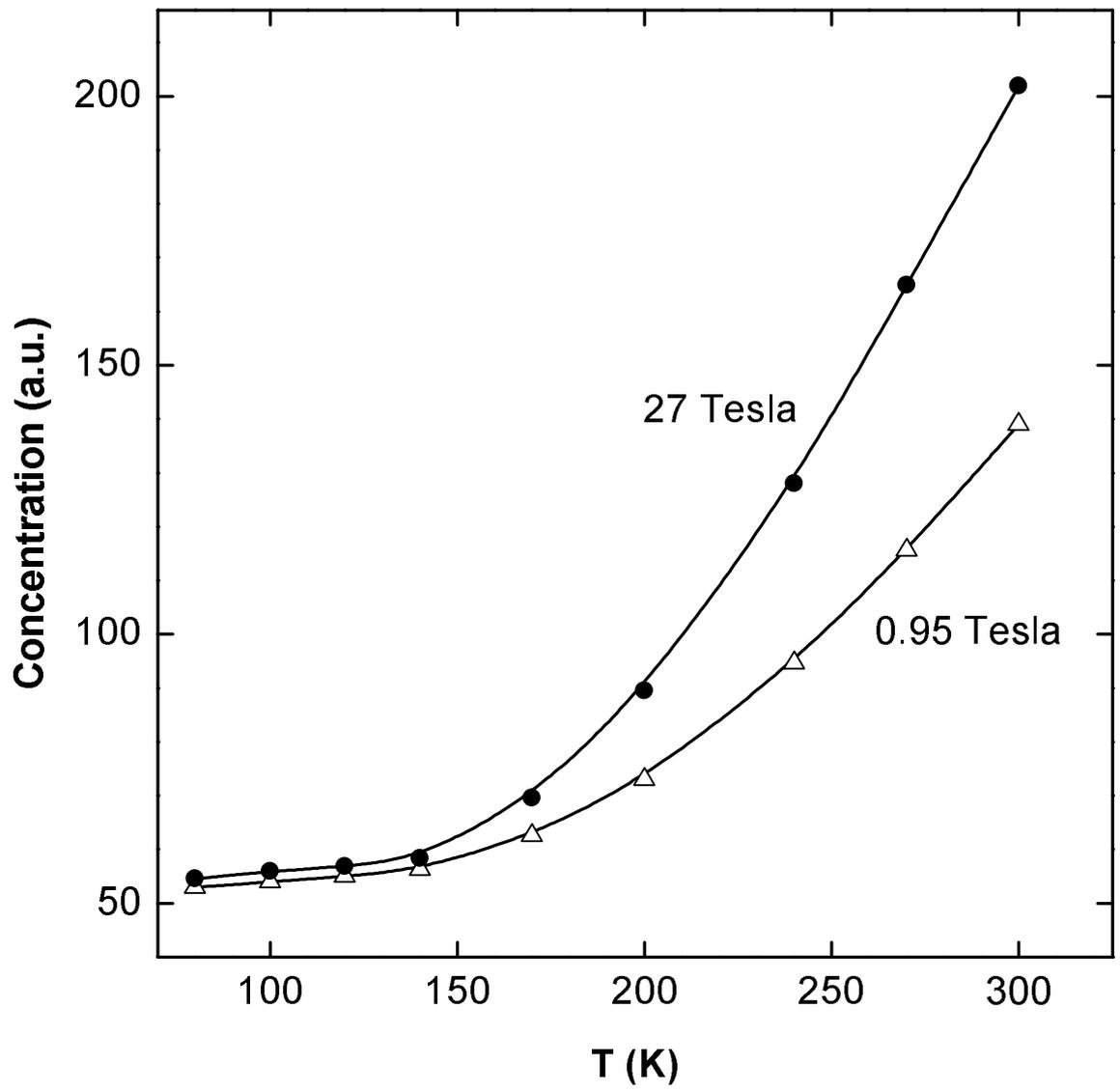
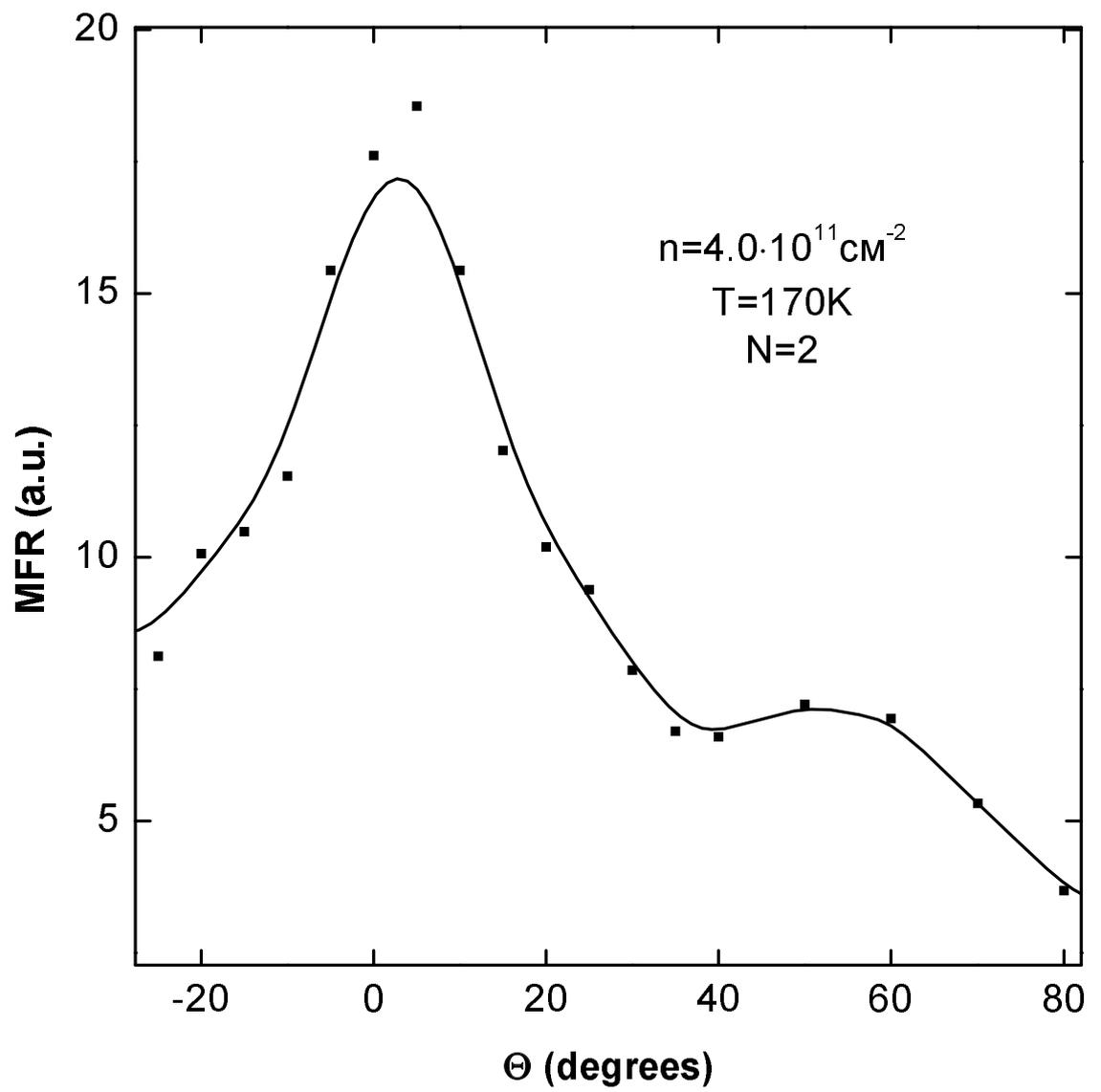
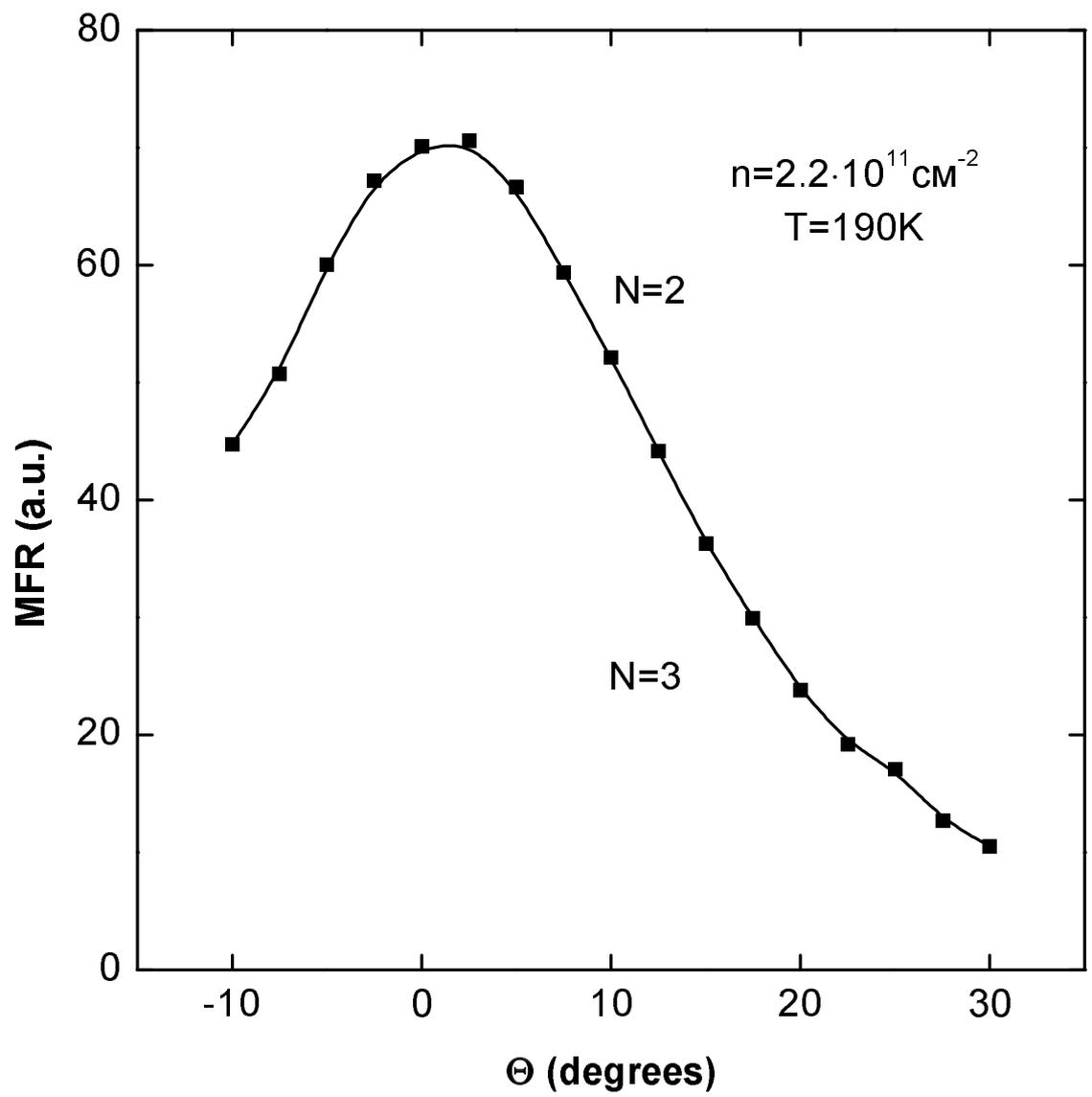
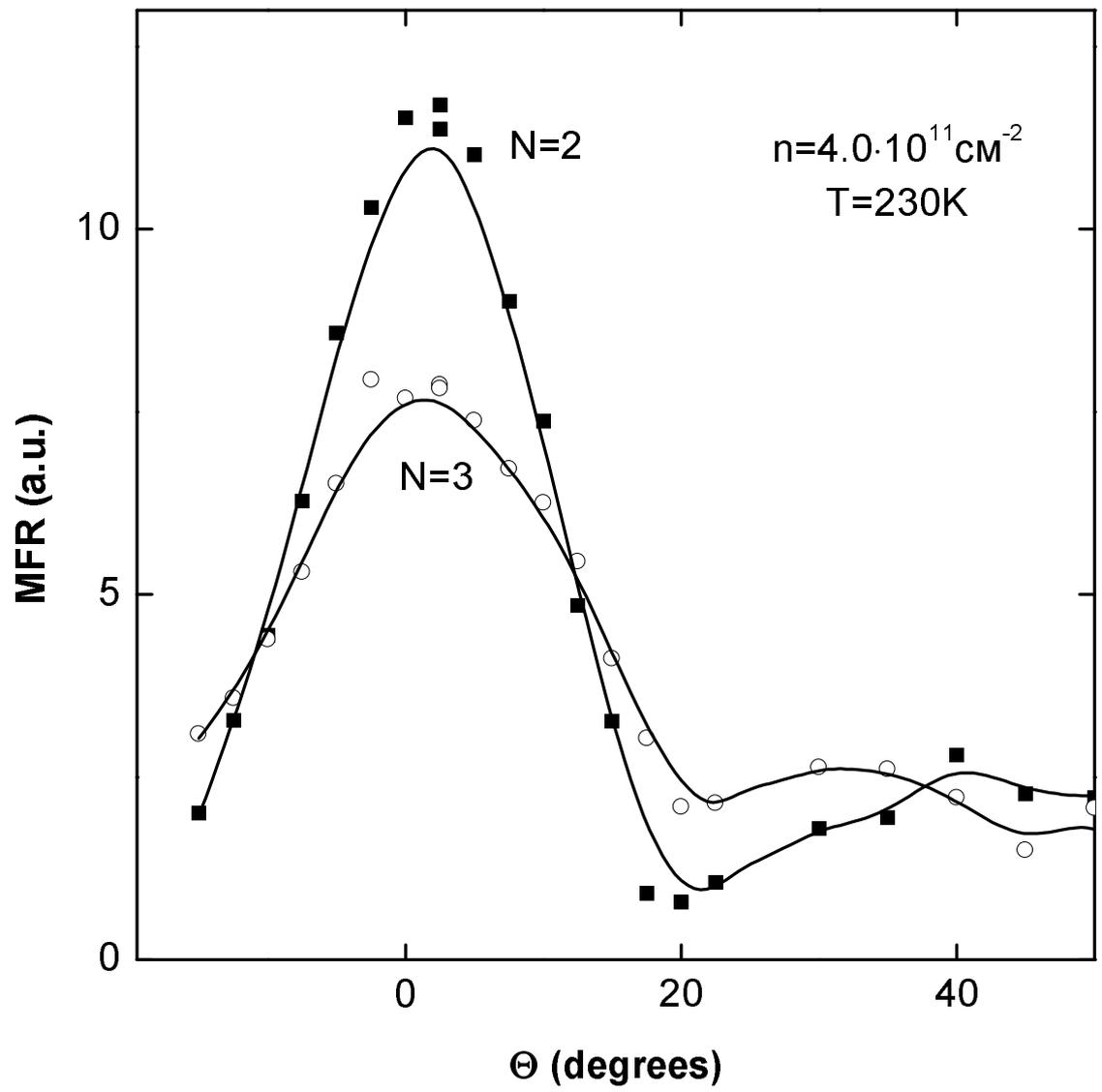
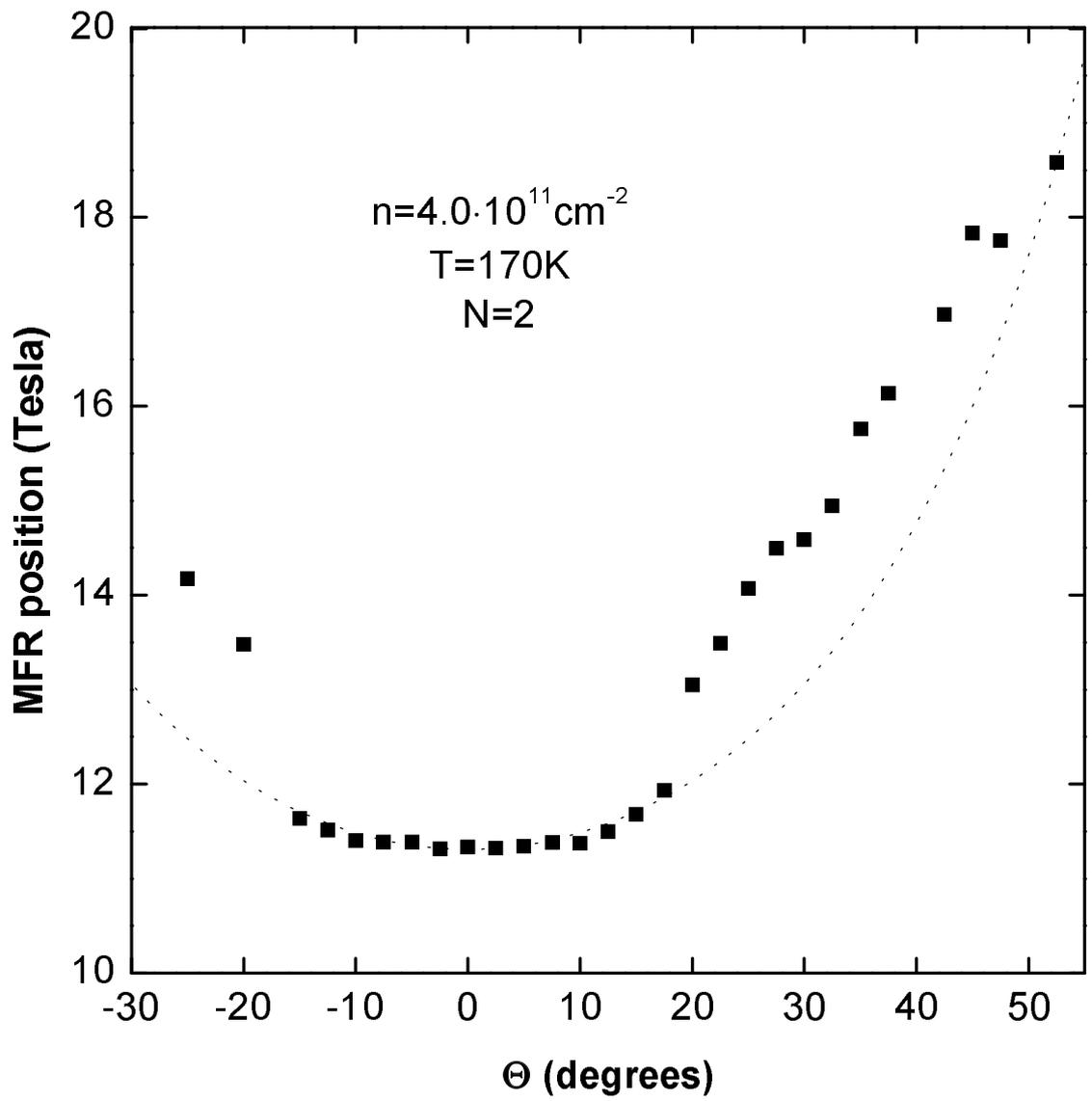
\end{document}